\documentclass[a4paper,12pt,aip,pop,superscriptaddress,reprint,floatfix]{revtex4-1}

\usepackage[utf8]{inputenc}
\usepackage[english]{babel}
\usepackage[T1]{fontenc}
\usepackage{amsmath,amssymb}
\usepackage{graphicx}
\usepackage{braket}
\usepackage{amsfonts}
\usepackage{array}
\usepackage{lmodern}
\usepackage{esvect}
\usepackage{fancyhdr}
\usepackage{xcolor}
\usepackage{textcomp}
\usepackage[squaren,Gray]{SIunits}
\usepackage{natbib}
\usepackage{hyperref}
\usepackage[caption=false]{subfig}

\graphicspath{{Images/}}

\begin{document} 

\title{Plasma dynamics of a laser filamentation-guided spark}

\author{Guillaume Point}
\email{guillaume.point@onera.fr}
\affiliation{Laboratoire d'Optique Appliqu\'ee, ENSTA ParisTech, CNRS, Ecole Polytechnique, Universit\'e Paris-Saclay, 828 boulevard des Mar\'echaux, 91762 Palaiseau cedex, France}
\affiliation{ONERA-CP, Chemin de la Hunière et des Joncherettes, 91123 Palaiseau cedex, France}
\author{Leonid Arantchouk}
\affiliation{Laboratoire d'Optique Appliqu\'ee, ENSTA ParisTech, CNRS, Ecole Polytechnique, Universit\'e Paris-Saclay, 828 boulevard des Mar\'echaux, 91762 Palaiseau cedex, France}
\affiliation{Laboratoire de Physique des Plasmas, CNRS, Ecole Polytechnique, Universit\'e Paris-Saclay, route de Saclay, 91128 Palaiseau cedex, France}
\author{J\'er\^ome Carbonnel}
\affiliation{Laboratoire d'Optique Appliqu\'ee, ENSTA ParisTech, CNRS, Ecole Polytechnique, Universit\'e Paris-Saclay, 828 boulevard des Mar\'echaux, 91762 Palaiseau cedex, France}
\author{Andr\'e Mysyrowicz}
\affiliation{Laboratoire d'Optique Appliqu\'ee, ENSTA ParisTech, CNRS, Ecole Polytechnique, Universit\'e Paris-Saclay, 828 boulevard des Mar\'echaux, 91762 Palaiseau cedex, France}
\author{Aur\'elien Houard}
\email{aurelien.houard@ensta-paristech.fr}
\affiliation{Laboratoire d'Optique Appliqu\'ee, ENSTA ParisTech, CNRS, Ecole Polytechnique, Universit\'e Paris-Saclay, 828 boulevard des Mar\'echaux, 91762 Palaiseau cedex, France}

\begin{abstract}
We investigate experimentally the plasma dynamics of a centimeter-scale, laser filamentation-guided spark discharge. Using electrical and optical diagnostics to study monopolar discharges with varying current pulses we show that plasma decay is dominated by free electron recombination if the current decay time is shorter than the recombination characteristic time. In the opposite case, the plasma electron density closely follows the current evolution. We demonstrate that this criterion holds true in the case of damped AC sparks, and that alternative current is the best option to achieve a long plasma lifetime for a given peak current.
\end{abstract}

\maketitle

\section{Introduction}

The discovery of laser filamentation in air by Braun \textit{et al.} in 1995 \cite{Braun1995} resulted in a considerable interest for the potential applications of this spectacular propagation regime for ultrashort pulses, characterized by the beam maintaining a very high intensity over several Rayleigh lengths \cite{Couairon2007}. Among other things, it was soon demonstrated that filamentation can trigger and guide spark discharges along the path followed by the propagating laser pulse \cite{Zhao1995,Vidal2000,Tzortzakis2001}.

This effect can be explained by the laser pulse energy deposited in air during propagation, chiefly through photoionization \cite{Rosenthal2015,Point2015b}, which leads to the formation and subsequent hydrodynamic expansion of a hot air cylinder along the filament, leaving a central air channel with reduced density \cite{Plooster1970,Plooster1971,Tzortzakis2001,Lenzner2013,Cheng2013,Wahlstrand2014,Point2015,Point2016}. This results in the breakdown voltage of air being diminished along the pulse propagation direction, hence the triggering and guiding ability of laser filamentation on spark discharges. Previously, powerful nanosecond lasers had already been used to trigger electric discharges by significantly reducing the breakdown threshold of air \cite{Koopman1971,Guenther1978,Miki1996}. However in this case the laser energy is deposited in a very inhomogeneous way in the form of discrete bubbles of dense and hot plasma whereas filamentation acts continuously and, therefore, yields better results.

Laser-triggered discharges are currently investigated for the development of several pending applications. For example, this may provide easily deployable and reconfigurable radio-frequency (RF) antennas \cite{Dwyer1984,Brelet2012} as well as potentially act as a lightning rod \cite{Zhao1995,Comtois2003,Kasparian2008}. They can also be used for contactless electrical power transmission \cite{Klapas1976,Houard2007,Arantchouk2013}. All these applications require a precise control of plasma parameters, especially electron density. This can be achieved using standard plasma diagnostics such as interferometry or diffractometry \cite{Hutchinson2002}. However such diagnostics usually involves cumbersome experimental tools and data processing techniques.

In this Article we study centimeter-scale, filamentation-triggered sparks with $\sim \unit{100}{\ampere}$ peak current and $\sim \unit{1}{\micro\second}$ duration. The time evolution of the electron density is recorded by means of a two-color interferometer \cite{Point2014}, while a simple current viewing resistor is used to record the time trace of the discharge current. By varying the amplitude and duration of the current pulses in the monopolar regime we find that the discharge plasma decay follows two different regimes: if the discharge current evolves much faster than the plasma recombines, then the electron density damps following a hyperbola, characteristic of recombination. However if the discharge current evolves slowly, the electron density instantaneous value follows closely the current time evolution. We also show these results to be valid in the case of an AC discharge. Thanks to this simple criterion, it becomes possible to have a relatively precise estimate of the instantaneous electron density in the discharge plasma using electrical diagnostics only. It also shows that the evolution of electron density can be easily controlled using a slowly varying current waveform.

\section{Experimental setup}

\begin{figure}[!ht]
\begin{center}
\includegraphics[width=.49\textwidth]{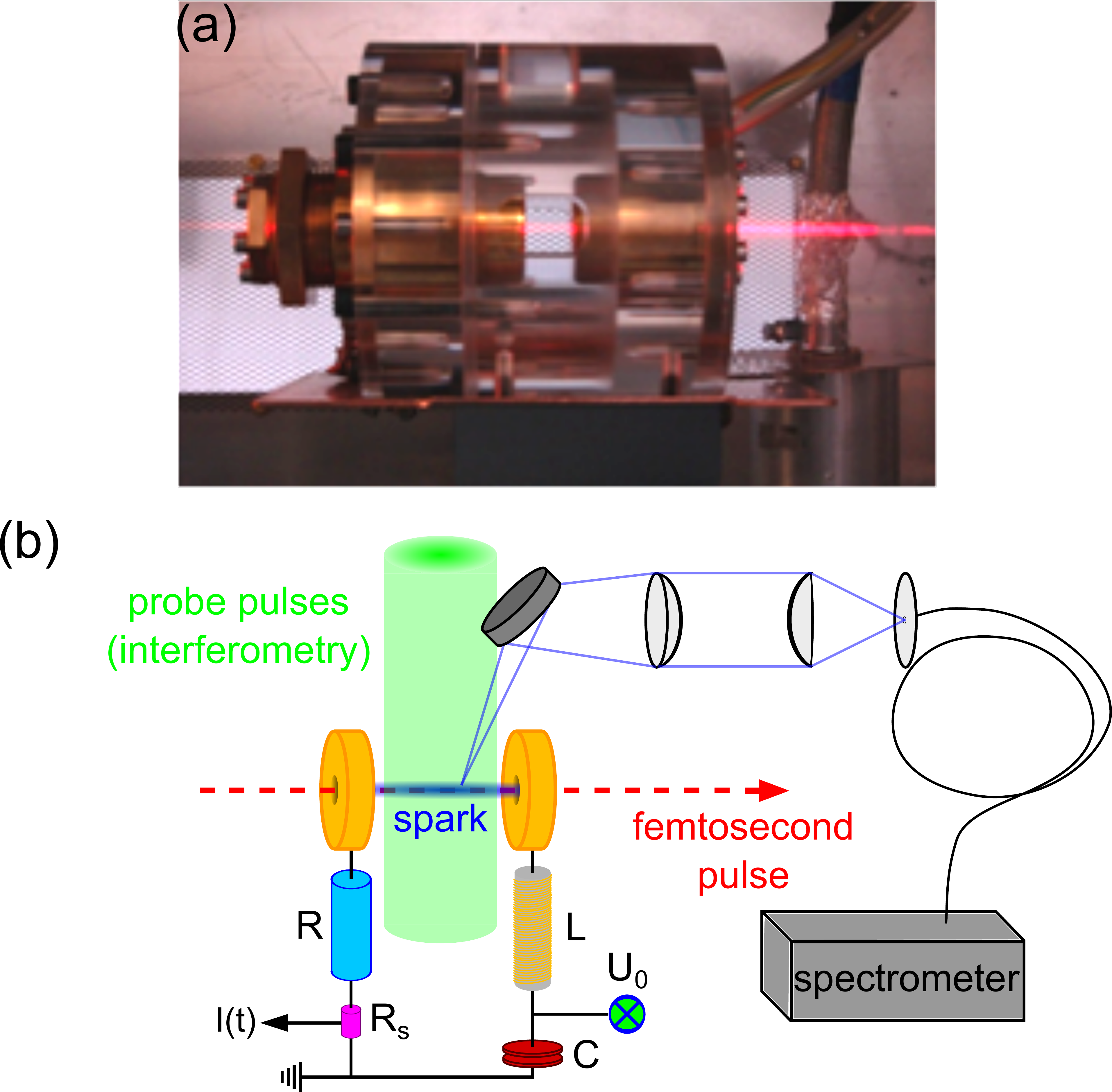}
\end{center}
\caption{Experimental setup for the study of centimeter-scale sparks. (a): photograph of the spark gap used in the experiments. (b): schematic description of the experimental setup, displaying the electric circuit and the different diagnostics used: a current viewing resistor R$_s$, transverse two-color interferometry on the plasma and plasma luminescence spectroscopic analysis.}
\label{figure_1}
\end{figure}

The first part of this Article deals with a study of the evolution of the plasma from centimeter-scale, filamentation-guided spark discharges. To this purpose, we used the same gap switch as used by Arantchouk and co-authors \cite{Arantchouk2013} (see figure \ref{figure_1}-(a)). It is made of two cylindrical electrodes separated by a \unit{1}{\centi\metre} gap drilled with \unit{3}{\milli\metre} holes through which a laser pulse can propagate.

The corresponding electrical circuit is simple. It consists of a series RLC circuit, in which the capacitor C$= \unit{2}{\nano\farad}$ is charged to a voltage U$_0 = \unit{15}{\kilo\volt}$ DC (figure \ref{figure_1}-(b)). This voltage is well below the self-breakdown threshold of the switch U$_{cr} = \unit{30.7}{\kilo\volt}$, ensuring that no discharge can occur without external triggering. When a laser filament is formed in the gap, air breakdown voltage decreases down to U$_0$ after a delay on the order of \unit{10}{\nano\second} \cite{Tzortzakis2001}, resulting in the formation of a plasma channel closing the circuit and enabling the charge stored in C to flow.

The discharge evolution is monitored by various diagnostics. First, a current viewing resistor R$_s = \unit{50}{\milli\ohm}$ with a \unit{2}{\giga\hertz} bandwidth enables us to closely follow the time evolution of the discharge current. Second, electron density and neutral density profiles in the discharge are recorded in the single shot regime by means of the two-color interferometric system described in reference \cite{Point2014}. This instrument makes use of both the first and second harmonic from a Nd:YAG probe laser and has a \unit{8}{\nano\second} temporal and a \unit{10}{\micro\metre} spatial resolution with a field of view covering the whole discharge gap. Its sensitivity is estimated to be around $\unit{10^{22}}{\rpcubic\metre}$ for electron and $\unit{10^{24}}{\rpcubic\metre}$ for neutral densities. Finally, plasma luminescence is collected using an inverted telescope and analyzed by a spectrometer (model USB 2000+ from Ocean Optics), yielding time-integrated emission spectra.

\section{Influence of the current waveform on the plasma in the monopolar regime}

\begin{figure}[!ht]
\begin{center}
\includegraphics[width=.45\textwidth]{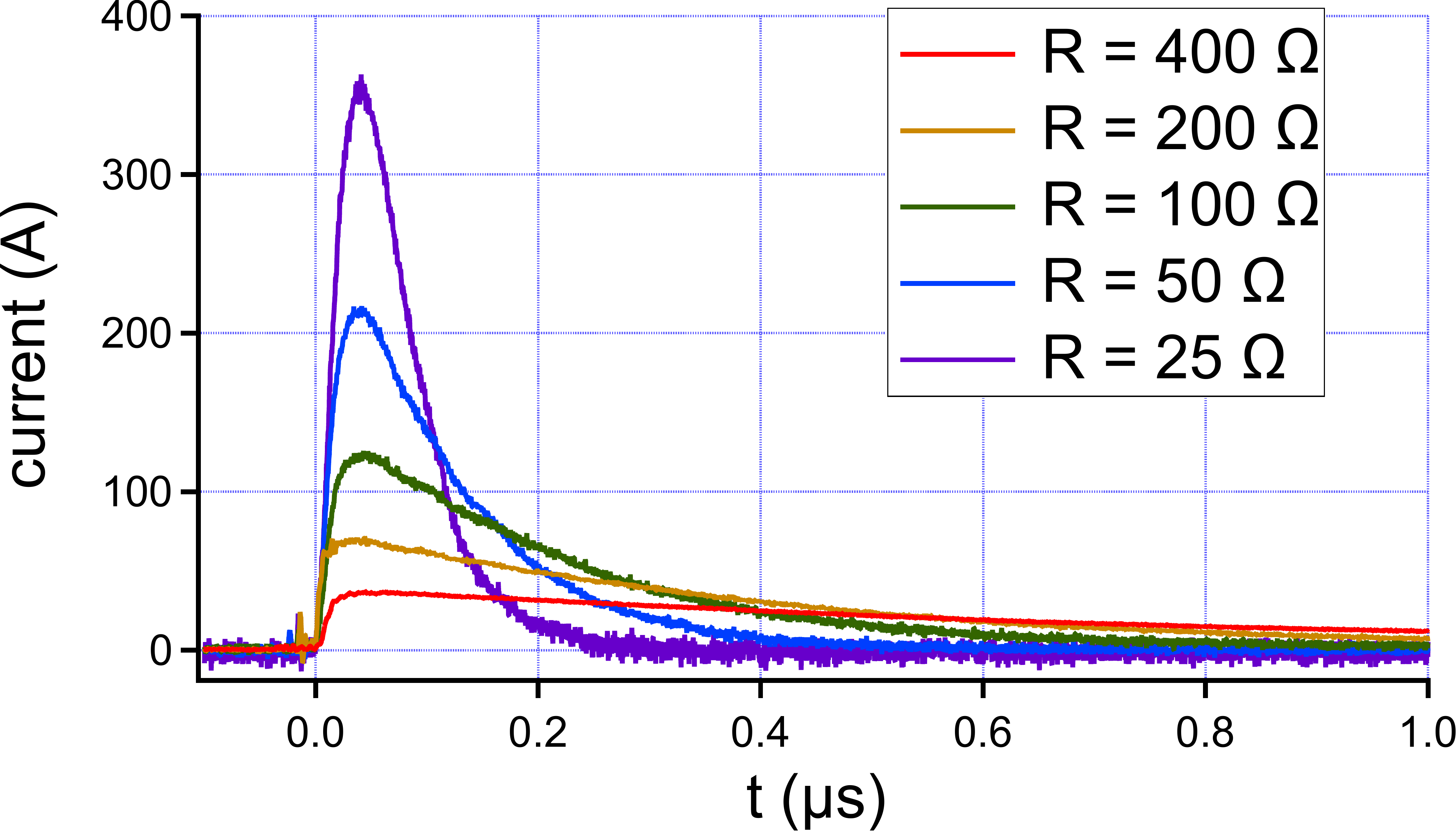}
\end{center}
\caption{Monopolar current pulses generated during the study.}
\label{figure_2}
\end{figure}

The first performed study was to investigate the influence of the discharge current waveform on the evolution of the spark plasma in the monopolar regime. In this case, the charge in the capacitor C is kept constant while the ballast resistor R is varied between 25 and \unit{400}{\ohm}. As the circuit self-inductance L is very low (L $\approx \unit{1}{\micro\henry}$), current always evolves in the overdamped regime, characterized by an exponential decay corresponding to a RC discharge. Corresponding current pulses have an amplitude ranging from 36 to \unit{352}{\ampere} and a duration varying from \unit{4}{\micro\second} to \unit{100}{\nano\second} (figure \ref{figure_2}).

\begin{figure}[!ht]
\begin{center}
\includegraphics[width=.48\textwidth]{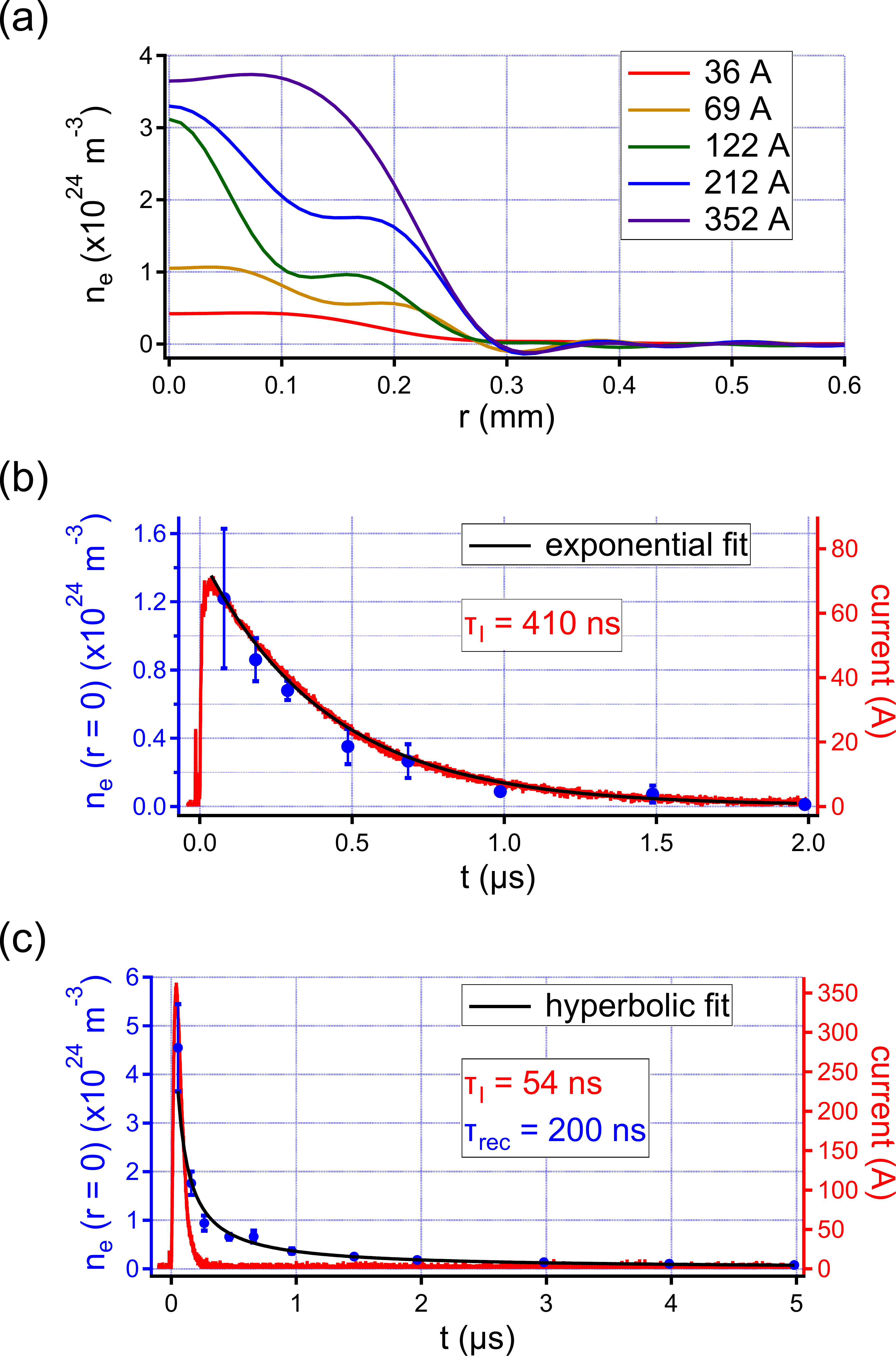}
\end{center}
\caption{(a): radial electron density profiles for the different current waveforms recorded at delay \unit{50}{\nano\second} after the discharge onset, midway of the plasma channel. (b): temporal evolution of the on-axis electron density for the \unit{69}{\ampere} current pulse. Error bars correspond to $\pm$ 1 standard deviation over a 5 shot statistics. Electron density evolves according to current, following an exponential damping. (c): temporal evolution of the on-axis electron density for the \unit{352}{\ampere} current pulse. Error bars correspond to $\pm$ 1 standard deviation over a 5 shot statistics. Electron density evolution is now dominated by recombination, as evidenced by the hyperbolic fit.}
\label{figure_3}
\end{figure}

Two-color interferometry enables us to record space and time-resolved electron density radial profiles from discharge plasma, assuming a local cylindrical symmetry. Figure \ref{figure_3}-(a) displays examples of such profiles recorded at an early time. These profiles are characterized by a central electron peak, amplitude and width of which are well ordered with respect to peak current, that is the more intense the pulse, the denser and wider the electron density profile. Recorded peak densities lie around $\unit{10^{24}}{\rpcubic\metre}$, that is an ionization rate of a few percents.

Studying the decay of electron density in time, it was found that there are two different regimes depending on the current pulses, with a transition lying around a \unit{100}{\ampere} peak current. As seen in figure \ref{figure_3}-(b), for low amplitude and long current pulses with a relatively smooth current time derivative, plasma damping is dictated by current evolution, in this case following the same exponential trend. 

When the peak current is increased and the pulse duration shortened, this is no longer true (figure \ref{figure_3}-(c)). The temporal evolution of electron density is instead well fitted by a hyberbola. This is a mark of plasma recombination because the evolution equation of electron density in an isolated plasma where diffusion processes can be neglected is given by:
\begin{equation}
\partial_t n_e(\vv{r},t) = -\beta_{eff}n_e(\vv{r},t)^2,
\end{equation}
where $\beta_{eff}$ is an effective recombination coefficient taking all recombination processes into account \cite{Point2015b}. The solution to this equation is given by:
\begin{equation}
n_e(\vv{r},t) = \frac{n_e(\vv{r},0)}{1+\beta_{eff}n_e(\vv{r},0)t},
\end{equation}
where $n_e(\vv{r},0)$ is the initial electron density. We define the corresponding characteristic recombination time $\tau_{rec}$ by:
\begin{equation}
\tau_{rec} = \frac{2}{\beta_{eff}n_e(\vv{r},0)}.
\end{equation}

Using this definition, $\tau_{rec}$ is a $1/3$ characteristic decay time, close to the widely used $1/\mathrm{e}$ characteristic decay time. By fitting the time evolution of the electron density for the high current discharges, in which case plasma density does not follow the current evolution, we could extract the recombination time $\tau_{rec}$, yielding the following results:
\begin{equation}
\left\{
\begin{array}{lll}
\tau_{rec}(\unit{122}{\ampere}) = \unit{162 \pm 34}{\nano\second}\\
\tau_{rec}(\unit{212}{\ampere}) = \unit{162 \pm 46}{\nano\second}\\
\tau_{rec}(\unit{352}{\ampere}) = \unit{206 \pm 42}{\nano\second}.
\end{array}
\right.
\label{eq_recomb_rate}
\end{equation}
It appears that $\tau_{rec}$ remains almost constant as the peak current is increased.

We explain the regime transition for plasma decay by comparing $\tau_{rec}$ to the characteristic time of the current evolution, $\tau_{I}$. If $\tau_{rec} < \tau_I$, then the current dictates the evolution of electron density because the plasma has time to adapt to current changes in terms of density and temperature. However when $\tau_{rec} \gtrsim \tau_I$, the current varies too quickly and the plasma is unable to follow this rapid evolution. If we go back to our example cases, the recombination time for the \unit{69}{\ampere} current pulse can be supposed to be similar to that of higher currents, since this parameter remains almost constant, that is $\tau_{rec} \approx \unit{160}{\nano\second}$. As for the current characteristic time, it is given by the circuit time constant $\mathrm{RC} = \unit{400}{\nano\second}$. We therefore have $\tau_I > \tau_{rec}$ and the electron density evolution depends on the current evolution. In the case of the \unit{352}{\ampere} current pulse $\tau_I = \unit{54}{\nano\second}$, which is much lower than $\tau_{rec}$, and the time evolution of the electron density is dominated by recombination.

\begin{figure}[!ht]
\begin{center}
\includegraphics[width=.48\textwidth]{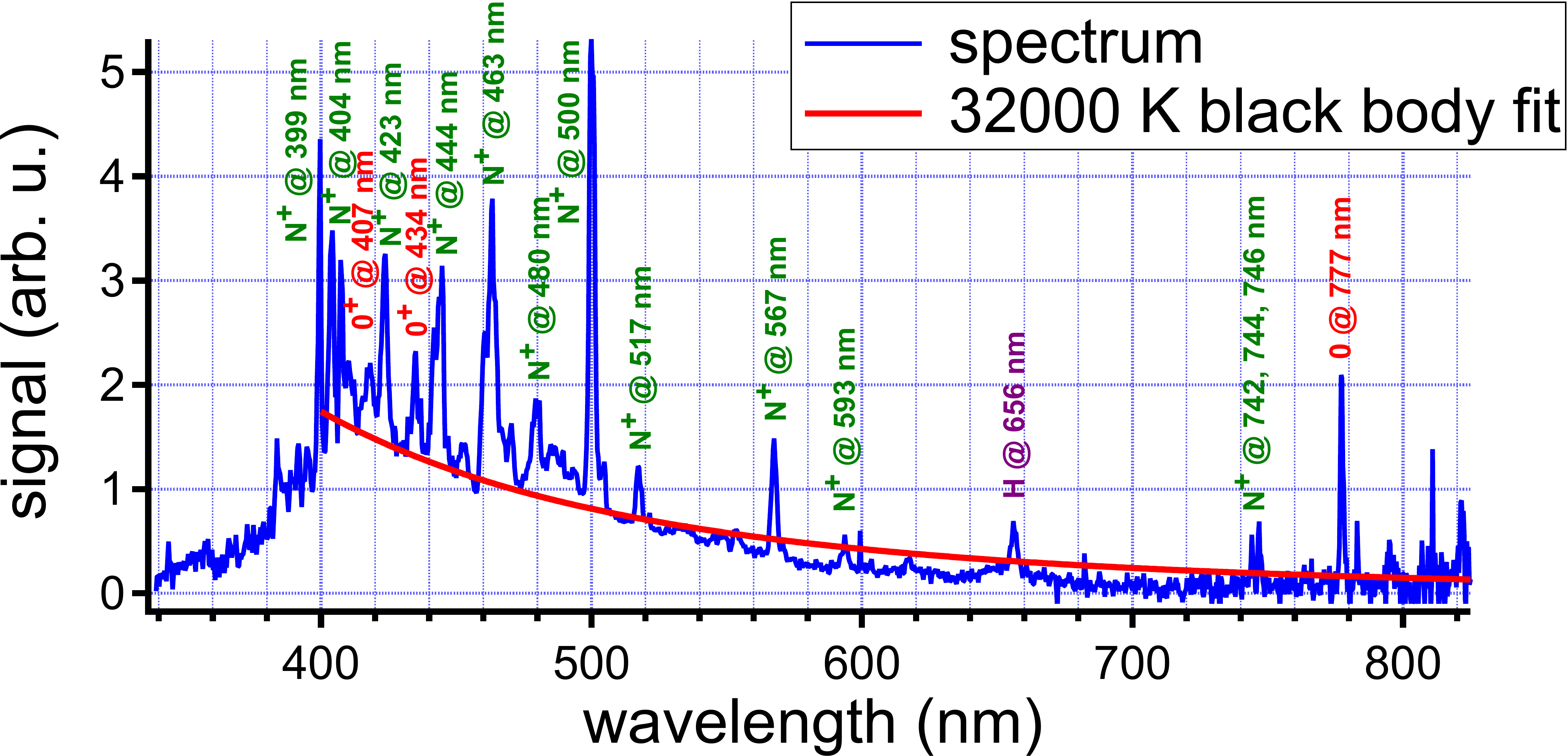}
\end{center}
\caption{Time-integrated emission spectrum of the discharge plasma recorded for the \unit{36}{\ampere} current pulse, corrected for the spectral response of the optical system. Main emission lines are indicated on the graph: green for N$^+$ lines, red for O and O$^+$ lines and violet for H lines.}
\label{figure_4}
\end{figure}

We also recorded the time-integrated emission spectrum of the discharge plasma to further investigate its properties. The spectrum obtained with the \unit{36}{\ampere} pulse is plotted in figure \ref{figure_4}. This spectrum is characteristic of a thermal plasma with a continuum component and emission lines superimposed over it. Almost all identified lines correspond to singly-ionized atomic nitrogen and oxygen, with the notable exception of the hydrogen Balmer-$\alpha$ emission line at \unit{656}{\nano\metre} and of the atomic oxygen triplet at \unit{777}{\nano\metre}. As seen on the spectrum, the continuum component is rather well fitted by a black body emission curve, yielding an estimated plasma temperature of $\sim \unit{3}{\electronvolt}$. Discrepancy below \unit{400}{\nano\metre} is due to the response function of our light collection system, which collapses quickly in the near-UV and renders recordings unreliable. The coexistence of emission lines with a black body behavior is explained by the fact that the spectrum is time-integrated. Black body contribution comes from early times when the plasma is at its densest and hottest while it becomes optically thin later on, resulting in the numerous recorded atomic emission lines. In means that the estimated plasma temperature is the \textit{initial} plasma temperature. A \unit{30}{\kilo\kelvin} temperature level is indeed typical for spark discharge events \cite{Plooster1971,Shneider2006}.

\section{Comparison with the AC regime}

We now compare the plasma evolution observed with monopolar current pulses with that measured with an alternating current (AC). To this purpose a solenoid coil with a \unit{28}{\micro\henry} inductance is introduced in the electric circuit described in figure \ref{figure_1}-(b), resulting in the discharge current following  exponentially-damped sinusoidal oscillations. 

\begin{figure}[!ht]
\begin{center}
\includegraphics[width=.48\textwidth]{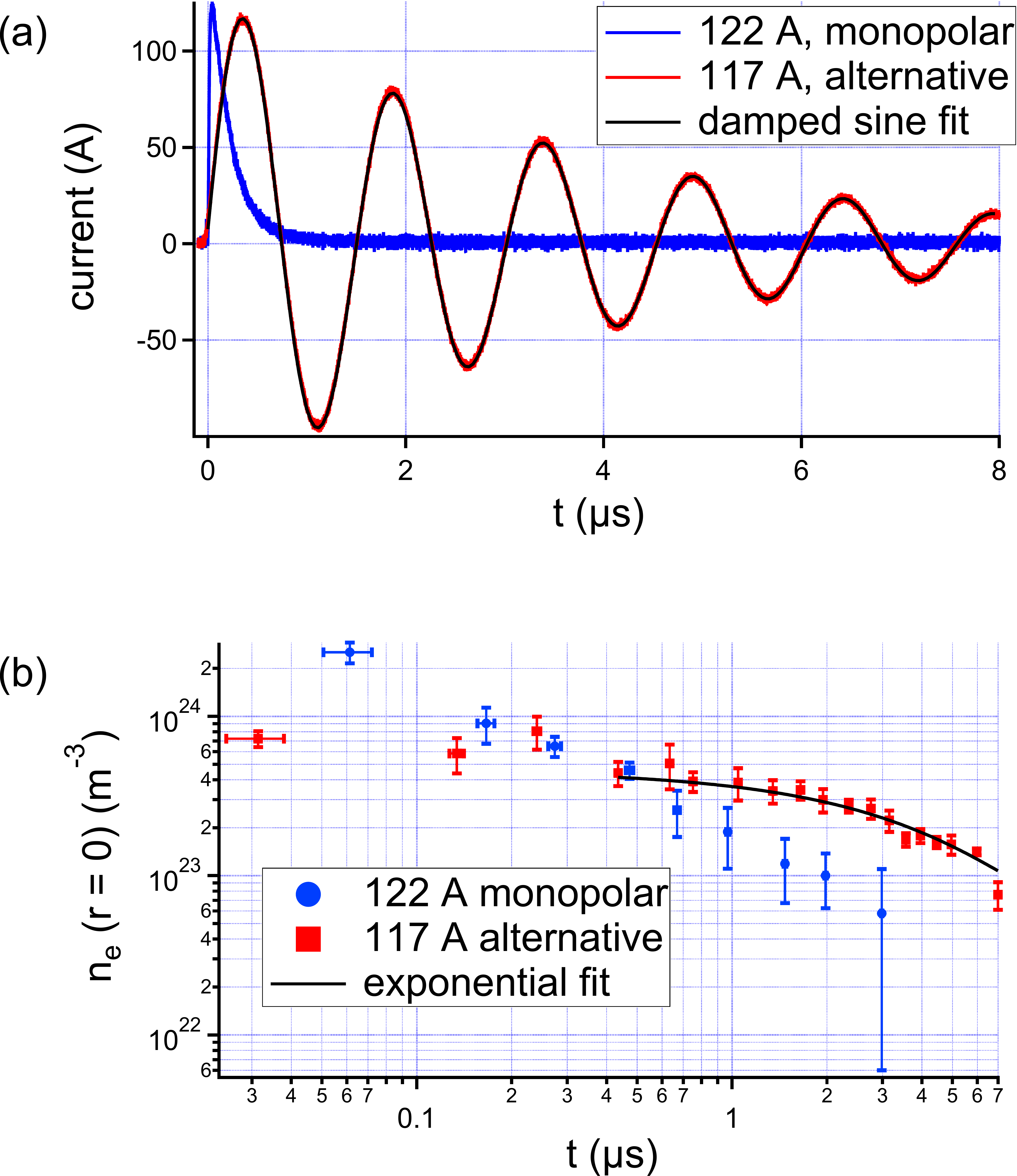}
\end{center}
\caption{(a): current waveform measured in the AC regime (red curve) with corresponding exponentially-damped sine fit (black curve), and monopolar current pulse with similar current amplitude (blue curve). (b): time evolution of on-axis electron density in the AC regime (red squares) with an exponential fit (black curve), or in the monopolar regime (blue circles).}
\label{figure_5}
\end{figure}

A curve fitting (figure \ref{figure_5}-(a)) enabled us to retrieve the characteristic damping time $\tau$ and frequency $f_0$ of the waveform:
\begin{equation}
\left\{
\begin{array}{ll}
\tau = \unit{3.77}{\micro\second}\\
f_0 = \unit{659}{\kilo\hertz}.
\end{array}
\right.
\end{equation}

Since the reached peak current is \unit{117}{\ampere}, these discharges can be directly compared to monopolar discharges studied in the monopolar regime with a \unit{122}{\ampere} maximum current. The evolution of on-axis electron density for the two configurations is plotted in figure \ref{figure_5}-(b).

As seen in this last figure, electron density damps considerably more slowly in the AC regime than for the monopolar current pulse, remaining in a quasi-steady state for several microseconds. Looking more precisely at the decreasing trend for $n_e$, we can see that its evolution is well fitted by an exponential decay, the characteristic time of which is equal to that of the current damping time. Indeed in this case there are two different current characteristic times. The first one corresponds to the fast sinusoidal oscillations and is given by:
\begin{equation}
\tau_{I,1} = \frac{1}{4\pi f_0} = \frac{1}{2\omega_0} = \unit{120}{\nano\second}.
\end{equation}
Here the factor $1/2$ was introduced because the plasma reacts to the absolute value of the current, yielding a period equal to half the period of signed current oscillations. The second current characteristic time corresponds to the slow exponential decay of the waveform and is equal to $\tau_{I,2} = \tau = \unit{3.77}{\micro\second}$. If we assume the recombination rate in the AC regime to be the same as in the monopolar regime, we can expect to have $\tau_{rec} \approx \unit{200}{\nano\second}$. We thus have $\tau_{rec} > \tau_{I,1}$ while $\tau_{rec} \ll \tau_{I,2}$. According to the criterion established in the previous section, electron density would therefore follow the long-time exponential current decrease, but not the fast current oscillations, which is precisely what is seen in figure \ref{figure_5}.

\section{Conclusions}

In this Article, we devised a simple criterion on the time evolution of electron density in a laser-induced spark discharge plasma: if the characteristic recombination time is higher than the discharge current evolution, then electron density decays following a hyperbola, mark of recombination. Conversely, if recombination is faster than the current characteristic evolution time, the electron density will be pegged to the discharge current decay. This criterion provides both a quick way to estimate electron density in a discharge plasma, based on a measurement of the discharge current only, and shows that electron density can be easily controlled by using an appropriate current waveform. This criterion was shown to be valid both for monopolar and AC current waveforms.

\begin{acknowledgments}

This research work was funded by the French Direction G\'en\'erale de l'Armement (grant n$^{\mathrm{o}}$ 2013.95.0901).

\end{acknowledgments}

\bibliographystyle{apsrev4-1}
\bibliography{biblio}

\end{document}